\documentclass[aps,prd,nofootinbib,twocolumn]{revtex4}
\pdfoutput=1

\usepackage{graphicx}
\usepackage{hyperref}
\usepackage{color}

\begin{document}
\leftline{FTUV-16-01-29, IFIC/16-03}

\title{Eternal Hilltop Inflation}

\author{Gabriela Barenboim} 
\email{Gabriela.Barenboim@uv.es}
\affiliation{Departament de F\'isica Te\`orica and IFIC, Universitat de Val\`encia-CSIC, E-46100, Burjassot, Spain}

\author{William H.\ Kinney} 
\email{whkinney@buffalo.edu}
\affiliation{Dept. of Physics, University at Buffalo,
        239 Fronczak Hall, Buffalo, NY 14260-1500}

\author{Wan-Il Park} 
\email{Wanil.Park@uv.es}
\affiliation{Departament de F\'isica Te\`orica and IFIC, Universitat de Val\`encia-CSIC, E-46100, Burjassot, Spain}

\date{\today}

\begin{abstract}
We consider eternal inflation in hilltop-type inflation models, favored by current data, in which the scalar field in inflation rolls off of a local maximum of the potential. Unlike chaotic or plateau-type inflation models, in hilltop inflation the region of field space which supports eternal inflation is finite, and the expansion rate $H_{EI}$ during eternal inflation is almost exactly the same as the expansion rate $H_*$ during slow roll inflation. Therefore, in any given Hubble volume, there is a finite and calculable expectation value for the lifetime of the ``eternal'' inflation phase, during which quantum flucutations dominate over classical field evolution. We show that despite this, inflation in hilltop models is nonetheless eternal in the sense that the volume of the spacetime at any finite time is exponentially dominated by regions which continue to inflate. This is true regardless of the energy scale of inflation, and eternal inflation is supported for inflation at arbitrarily low energy scale. 
\end{abstract}

\maketitle

\section{Introduction}
\label{sec:Intro}

Inflation \cite{Starobinsky:1980te,Sato:1981ds,Sato:1980yn,Kazanas:1980tx,Guth:1980zm,Linde:1981mu,Albrecht:1982wi} is now well-established as the leading theory of the very early universe. In addition to explaining the flatness and homogeneity of the cosmos, inflation predicts the generation of perturbations from quantum fluctuations in the early universe \cite{Starobinsky:1979ty,Mukhanov:1981xt,Mukhanov:2003xw,Linde:1983gd,Hawking:1982cz,Hawking:1982my,Starobinsky:1982ee,Guth:1982ec,Bardeen:1983qw}, a prediction which has been tested to high precision in measurements of the Cosmic Microwave Background (CMB), most recently by the Planck satellite \cite{Ade:2015lrj}, and the BICEP/Keck array \cite{Ade:2015tva,Array:2015xqh}

Typically, inflation models have the property of being {\it eternal}. This is most easily illustrated in the case of simple chaotic inflationary models, characterized by potentials of the form $V\left(\phi\right) \propto \phi^p$, which have field excursion $\Delta\phi > M_{\rm P}$ during inflation, where $M_{\rm P} = m_{\rm P} / \sqrt{8 \pi} \sim 2 \times 10^{18}\ {\rm GeV}$ is the reduced Planck Mass. Such potentials have the property that, for field values $\phi \gg M_{\rm P}$, the amplitude of quantum fluctuations in the field becomes larger than the classical field variation, so that the field is as likely to roll {\it up} the potential as it is to roll down the potential. Therefore, in a statistical sense, inflation never ends: there will always be regions of the universe where the field has fluctuated upward, rather than downward, and inflation becomes a quasi-stationary, infinitely self-reproducing state of eternal inflation \cite{Vilenkin:1983xq,Guth:1985ya,Linde:1986fc,Linde:1986fd,Starobinsky:1986fx}. However, such simple chaotic potentials are now ruled out by CMB data. Data now strongly favor ``hilltop'' or ``plateau''-type inflationary models \cite{Kohri:2007gq,Martin:2013tda,Barenboim:2013wra,Coone:2015fha,Huang:2015cke,Vennin:2015egh}, in which the inflaton rolls off a local maximum of the potential. Furthermore, an upper limit on the ratio $r$ of tensor to scalar perturbations of $r < 0.07$ \cite{Array:2015xqh} imposes a corresponding upper bound on the energy density during inflation,
\begin{equation}
V_0^{1/4} \simeq \left(3.3 \times 10^{16}\ {\rm GeV}\right) \times r^{1/4} < 1.7 \times 10^{16}\ {\rm GeV}.
\end{equation}
There is no corresponding lower bound on the inflationary energy scale, and inflation at very low energy scales is a good fit to data. 

In this paper, we consider eternal inflation in generic hiltop-type inflation models, and show that such potentials support eternal inflation at arbitrarily low energy scale. The paper is structured as follows: In Sec. \ref{sec:EternalInflation}, we discuss eternal inflation in a general sense, and derive bounds on eternal inflation in representative chaotic inflation models. In Sec. \ref{sec:TopHat}, we consider a simplified toy model of hilltop inflation, and derive bounds on parameters necessary for eternal inflation to occur. In Sec. \ref{sec:Hilltop}, we apply the result to generic hilltop inflation models. Section \ref{sec:Conclusions} presents discussion and conclusions.  

\section{Eternal Inflation}
\label{sec:EternalInflation}

Inflation is generically defined as a period of accelerating cosmological expansion, driven by a scalar field $\phi$ (the {\it inflaton}) with a potential which dominates over the kinetic energy of the field,
\begin{equation}
\rho_\phi = \frac{1}{2} \dot\phi^2 + V\left(\phi\right) \simeq V\left(\phi\right).
\end{equation}
In such a case, the universe expands quasi-exponentially. 
So-called ``eternal'' inflation occurs when quantum fluctuations in the inflaton field dominate over the classical field evolution: the amplitude of quantum fluctuations in the inflaton field during inflation is
\begin{equation}
\left\langle\delta\phi\right\rangle_Q \equiv \left\langle \delta \phi^2\right\rangle^{1/2} = \frac{H}{2 \pi}.
\end{equation}
Here we take a flat, Friedmann-Robertson-Walker universe with metric
\begin{equation}
ds^2 = dt^2 - a^2(t) d{\bf x}^2,
\end{equation}
where the expansion rate is $H \equiv \dot a / a$. The field evolution, including quantum fluctuations, is then stochastic and is governed by a Langevin Equation,
\begin{equation}
\ddot\phi + 3 H \dot\phi + \frac{d V}{d \phi} = N\left(t,{\bf x}\right),
\end{equation}
where $N\left(t,{\bf x}\right)$ is a Gaussian noise function generated by the quantum fluctuations \cite{Starobinsky:1986fx}.For eternal inflation to occur, this quantum fluctuation amplitude must be larger than the classical field variation over approximately a Hubble time,
\begin{equation}
\delta\phi_{Cl} = \frac{\dot\phi}{H}.
\end{equation}
Therefore, a simple condition for eternal inflation can be written
\begin{equation}
\frac{\left\langle\delta\phi\right\rangle_Q}{\delta\phi_{Cl}} = \frac{H^2}{2 \pi \dot\phi} > 1.\label{eq:eternal}
\end{equation}
We note that the fraction (\ref{eq:eternal}) is identical to the amplitude of the curvature perturbation for modes crossing the horizon during inflation,
\begin{equation}
P\left(k\right) = \frac{H^2}{2 \pi \dot\phi}.
\end{equation}
The curvature perturbation is simply the amplitude of quantum fluctuations in the inflaton in units of the field variation in a Hubble time! Accordingly, the condition for eternal inflation is that the curvature perturbation amplitude must exceed unity \cite{Goncharov:1987ir,Guth:2007ng},
\begin{equation}
P\left(k\right) > 1.
\end{equation}
We emphasize that, while this is a {\it necessary} condition, it is not itself a {\it sufficient} condition for eternal inflation \cite{Vennin:2015hra}, which we discuss below.
The simplest formal case we can consider is an exact power-law power spectrum,
\begin{equation}
P\left(k\right) = P\left(k_*\right) \left(\frac{k}{k_*}\right)^{n- 1},
\end{equation}
where we take $k_* = 0.05\ {\rm Mpc}^{-1}$, and, from Planck, $P\left(k_*\right) \simeq 2.2 \times 10^{-9}$ \cite{Ade:2015xua}. A power spectrum for which $n - 1 = {\rm const.}$ at all scales is generated by power-law background evolution of the form
\begin{equation}
a\left(t\right) \propto t^{1 / \epsilon},
\end{equation}
where $a\left(t\right)$ is the scale factor, and the {\it slow roll parameter} $\epsilon$ is determined by the equation of state:
\begin{equation}
\epsilon \equiv -\frac{a}{H} \frac{d H}{d a} = \frac{3}{2}\left(1 + \frac{p}{\rho}\right) = {\rm const.}
\end{equation}
The spectral index is related to $\epsilon$ as
\begin{equation} \label{eq:epsilonPL}
n - 1 = - 2 \epsilon.
\end{equation}
The curvature power spectrum for a comoving wavenumber $k$ is given by
\begin{equation} \label{eq:Pk}
P\left(k\right) = \left[\frac{H^2}{8 \pi^2 M_{\rm P}^2 \epsilon}\right]_{k = a H},
\end{equation}
where  the expression is evaluated when the perturbation modes exits the horizon, $k = a H$. Using the (\ref{eq:epsilonPL}), we can write the Hubble scale corresponding to $k_*$ as
\begin{equation}
\frac{H_*}{M_{\rm P}} = 2 \pi \sqrt{(1 - n) P\left(k_*\right)} \simeq 6 \times 10^{-5}. 
\end{equation}
The onset of eternal inflation occurs when wavenumber $k = k_{EI}$ is of order the Hubble length, and 
\begin{equation}
P\left(k_{EI}\right) = P\left(k_*\right) \left(\frac{k_{EI}}{k_*}\right)^{n - 1} = 1,
\end{equation}
or
\begin{equation}
\frac{k_{EI}}{k_*} = \left[P\left(k_*\right)\right]^{1 / \left(1 - n\right)} \sim 10^{-217},
\end{equation}
where we have taken $1 -n = 0.04$, consistent with the Planck best-fit. The order-unity curvature perturbations associated with eternal inflation only occur at scales exponentially larger than our current horizon size, and are therefore unobservable. We can calculate the Hubble parameter at the onset of eternal inflation from the relation
\begin{equation}
\frac{k}{H} \frac{d H}{d k} = \frac{\epsilon}{\epsilon - 1},
\end{equation}
so that 
\begin{equation} \label{eq:HEHSPL}
\frac{H_{EI}}{H_*} = \left(\frac{k_{EI}}{k_*}\right)^{\epsilon / \left(\epsilon - 1\right)} =  \left[P\left(k_*\right)\right]^{-1 / 2} \simeq 2.1 \times 10^{4}.
\end{equation}
We can then write the Hubble parameter $H_{EI}$ in Planck units,
\begin{equation} \label{eq:HEPL}
\frac{H_{EI}}{M_{\rm P}} = \left(\frac{H_{EI}}{H_*}\right) \left(\frac{H_*}{M_{\rm P}}\right) = 1.28,
\end{equation}
so that for a {\it pure power-law} power spectrum, eternal inflation only occurs when the expansion rate is of order the Planck scale, and quantum-gravitational effects are expected to be important. In this case, we have no sensible {\it semiclassical} description of eternal inflation. 

Power-law inflation, however, corresponds to a very particular choice of scalar field potential,
\begin{equation}
V\left(\phi\right) = V_0 e^{\lambda \left(\phi / M_{\rm P}\right)},
\end{equation}
and the relation (\ref{eq:HEPL}) is not generic to other choices of scalar field potential. Put in terms of the power spectrum, if the slope of the power spectrum changes with scale, eternal inflation can happen at much lower energy scale, or can be suppressed altogether \cite{Kinney:2014jya}. As an example of the former behavior, consider a simple linear scalar field potential,
\begin{equation} \label{eq:linearV}
V\left(\phi\right) = \Lambda^3 \phi.
\end{equation}
In the slow-roll limit,
\begin{equation}
\ddot \phi + 3 H \dot \phi \simeq 3 H \dot \phi = - V'\left(\phi\right),
\end{equation}
the slow-roll parameter $\epsilon$ is given by
\begin{equation}
\epsilon \simeq \frac{M_{\rm P}^2}{2} \left(\frac{V'\left(\phi\right)}{V\left(\phi\right)}\right)^2 = \frac{1}{2} \left(\frac{M_{\rm P}}{\phi}\right)^2.
\end{equation}
Inflation ends for $\phi < \phi_e$, where 
\begin{equation}
\epsilon\left(\phi_e\right) \equiv 1 \Rightarrow \phi_e = M_{\rm P} / \sqrt{2}.
\end{equation}
Defining the number of e-folds of expansion before the end of inflation as $dN = - d a / a$, 
\begin{equation}
N\left(\phi\right) = \frac{1}{M_{\rm P} \sqrt{2}} \int_{\phi_e}^{\phi}{\frac{d \phi}{\sqrt{\epsilon\left(\phi\right)}}} = \frac{1}{M_{\rm P}^2} \int_{\phi_e}^{\phi}{\phi d \phi},
\end{equation}
we have
\begin{equation}
\phi\left(N\right) = M_{\rm P} \sqrt{2 N + \frac{1}{2}}.
\end{equation}
Density perturbations of order the horizon size today were generated for $N = N_*$, where the exact number depends on the scale of inflation and thermal history after inflation \cite{Martin:2010kz,Munoz:2014eqa,Cai:2015soa,Cook:2015vqa}. For our purposes, a crude estimate of $N_* = [46,60]$ is sufficient. The curvature power spectrum (\ref{eq:Pk}) is given by
\begin{equation} \label{eq:Pklinear}
P\left(k_*\right) = \frac{4 N_* + 1}{8 \pi^2} \left(\frac{H_*}{M_{\rm P}}\right)^2 \simeq 2.2 \times 10^{-9}.
\end{equation}
Eternal inflation occurs at
\begin{equation} \label{eq:PElinear}
P(k_{EI}) = 1 = \frac{4 N_{EI} + 1}{4 N_* + 1} \left(\frac{H_{EI}}{H_*}\right)^2 P\left(k_*\right). 
\end{equation}
However, from the Friedmann Equation, the Hubble parameter is given in the slow roll limit by the potential (\ref{eq:linearV}),
\begin{equation}
\left(\frac{H}{M_{\rm P}}\right)^2 = \frac{V\left(\phi\right)}{3 M_{\rm P}^4} = \frac{1}{3} \left(\frac{\Lambda}{M_{\rm P}}\right)^3 \sqrt{2 N + \frac{1}{2}}.
\end{equation}
Therefore, we can write
\begin{equation}
\left(\frac{H_{EI}}{H_*}\right)^4 = \frac{4 N_{EI} + 1}{4 N_* + 1},
\end{equation}
and, from (\ref{eq:PElinear}),
\begin{equation}
\frac{H_{EI}}{H_*} = \left[P\left(k_*\right)\right]^{-1/6} \simeq 28,
\end{equation}
a much smaller figure than that for power-law inflation (\ref{eq:HEHSPL}). Taking $N_* = 60$ in  Eq. (\ref{eq:Pklinear}), we have
\begin{equation}
\frac{H_*}{M_{\rm P}} = \sqrt{\frac{8 \pi^2 P\left(k_*\right)}{4 N_* + 1}} = 2.7 \times 10^{-5},
\end{equation} 
so that $H_{EI}$ is much smaller than the Planck mass,
\begin{equation}
\frac{H_{EI}}{M_{\rm P}} \sim 10^{-3}.
\end{equation}
Eternal inflation therefore sets in well below the energy scale for which quantum gravity is relevant. This is typical of other chaotic inflation potentials, such as $V \propto \phi^2$ or $V \propto \phi^4$. However, eternal inflation in such models, where the potential (and therefore the Hubble parameter) is unbounded from above, is qualitatively different from hilltop inflation models, for which the field typically rolls away from a local maximum of the potential. We discuss such models in the next section. 

\section{A trivial hilltop model: The top-hat potential}
\label{sec:TopHat}

We first consider a simple ``top-hat'' potential to illustrate eternal inflation in hilltop-type inflation models. The potential is
\begin{equation}
V\left(\phi\right) = \left\lbrace{V_0 \ {\rm for} \ \left\vert\phi\right\vert \leq \Delta \atop 0 \ \ {\rm for} \ \left\vert\phi\right\vert > \Delta}\right\rbrace.
\end{equation}
Slow-roll inflation on such a potential is classically eternal, since the slow-roll attractor solution is:
\begin{equation}
\dot\phi = \frac{V'\left(\phi\right)}{3 H} = 0,
\end{equation}
and inflation continues forever. Such perfectly flat potentials have been studied previously as {\it ultra-slow roll} inflation \cite{Kinney:2005vj}, where it is seen that quantum fluctuations generate a scale-invariant spectrum of curvature fluctuations. 

However, while inflation continues forever classically, the field will also undergo quantum fluctuations. For simplicity and definiteness, we assume inflation begins with a homogeneous field $\phi_0 = \dot\phi_0 = 0$ on a manifold which we assume to be initially much larger than a Hubble volume. If the initial manifold is finite (for example, a toroid), probabilities are well-defined and we can treat the inflating space as a statistical ensemble of independently evolving Hubble volumes, and we avoid issues related to the well-known measure problem of eternal inflation \cite{Linde:1993nz,GarciaBellido:1994ci,Vanchurin:1999iv,Guth:2000ka,Easther:2005wi,Aguirre:2006ak,Aguirre:2006na,Bousso:2006ev,Winitzki:2006rn,Vilenkin:2006xv,Guth:2007ng,Linde:2007nm}. The question of {\it whether or not} eternal inflation occurs is independent of the question of how to impose a consistent measure on the eternally inflating spacetime itself, which will have infinite four-volume at timelike infinity and therefore still be subject to the measure problem.\footnote{We thank the referee for pointing this out.}
For field fluctuations following a Gaussian random distribution, the expectation value for the quantum fluctuations of the field in a Hubble time is just the expansion rate,
\begin{equation} \label{dphiQ}
\left\langle \delta\phi\right\rangle_Q = \frac{H}{2 \pi} = \frac{\sqrt{V_0}}{2 \pi \sqrt{3} M_{\rm P}}.
\end{equation}
We adopt a simple approximation to describe the stochastic evolution of the field. For a more precise treatment of stochastic field evolution, we refer the reader to, {\it e.g.}, \cite{Garriga:1997ef}. If the field executes a random walk the average lifetime $\left\langle t\right\rangle$ of the inflating state will be given by \cite{Starobinsky:1986fx}
\begin{equation}
\sqrt{H \left\langle t\right\rangle} = \frac{\Delta}{\left\langle \delta\phi\right\rangle_Q},
\end{equation} 
or
\begin{equation}
\left\langle t\right\rangle \simeq \frac{4 \pi^2 \Delta^2}{H^3}.
\end{equation}
Therefore, as a conservative approximation, the probability of inflation ending after time $t$ can be estimated for any Hubble patch generated from the original Hubble patch by inflation as \footnote{
More precisely, for the Gaussian random fluctuations of the field, the probability is given by
\begin{equation}
\Gamma_I(t) = {\rm erf} \left( \frac{\Delta}{\langle \delta \phi \rangle_Q \sqrt{2 Ht} } \right)
\end{equation}
In this case, $\Delta$ required for eternal inflation is smaller than our conservative estimate by a factor about $1.6$. } 
\begin{equation}
\Gamma_I\left(t\right) \propto e^{-t / \left\langle t\right\rangle} = \exp \left[- \left( \frac{\langle \delta \phi \rangle_Q}{\Delta} \right)^2 H t \right].
\end{equation}
However, during time $t$, an intial Hubble patch of initial volume $\mathcal V\left(0\right)$ will increase in volume  exponentially with time,
\begin{equation}
\Gamma_{\mathcal V} = \frac{{\mathcal V}\left(t\right)}{{\mathcal V}\left(0\right)} = e^{3 H t}.
\end{equation}
So the number of Hubble patches undergoing inflation at time $t$ is
\begin{equation}
\Gamma\left(t\right) \propto \exp{\left[\left(3 - \left( \frac{\langle \delta \phi \rangle_Q}{\Delta} \right)^2 \right) H t\right]}.
\end{equation}
If the exponent is positive, spacetime expansion dominates over the quantum fluctuations, and inflation can be said to be truly eternal. If the exponent is negative, inflation is exponentially quenched, and the manifold at late times is dominated by non-inflating regions. 

We therefore have a simple condition for the existence of eternal inflation,
\begin{equation} \label{eq:EIcondition}
\frac{\Delta}{\langle \delta \phi \rangle_Q} > \frac{1}{\sqrt{3}},
\end{equation}
or from Eq.~$(\ref{dphiQ})$, 
\begin{equation}
\Delta > \frac{\sqrt{V_0}}{6 \pi M_{\rm P}}.
\end{equation}
Therefore eternal inflation only occurs if the peak of the potential is sufficiently broad in units of its height $V_0$. This is qualitiatively different from the chaotic potentials considered in Sec. \ref{sec:EternalInflation}. For chaotic models, we generically have $H_{EI} \gg H_*$: eternal inflation occurs when the expansion rate becomes very large, so that $H^2 \gg \dot\phi$, and the curvature perturbation is of order unity. In the hilltop case, $H_{EI} \simeq H_*$, and the expansion rate during eternal inflation is nearly identical to that during the non-eternal phase of inflation. However, the classical field variation tends to zero near the maximum of the potential, $\dot\phi_{Cl} \propto V'\left(\phi\right) \rightarrow 0$, and the curvature perturbation $P\left(k\right) \propto H^2 / \dot\phi$ is again of order unity. In this sense, there are two distict classes of eternal inflation: models for which $H_{EI} \gg H_*$, and models for which $H_{EI} \sim H_*$. There is a second qualtitative difference between hilltop inflation and chaotic inflation: in the chaotic inflation case, eternal inflation will occur for arbitrarily large field fluctuations $\delta\phi$, since the field can always roll farther {\it up} the potential. Therefore, no relation of the type (\ref{eq:EIcondition}) exists.\footnote{Some authors have argued that inflation is not truly eternal even in this case. \cite{Gratton:2005bi,Kohli:2014ala}} This is also true of plateau-type inflation models, a simple example of which is a step function,
\begin{equation}
V\left(\phi\right) = \left\lbrace{V_0 \ {\rm for} \phi > 0 \atop 0 \ \ {\rm for} \ \phi <0}\right\rbrace.
\end{equation}
In this case, $H_{EI} \sim H_*$, but $\Delta \rightarrow \infty$. Starobinsky inflation \cite{Starobinsky:1980te,Kehagias:2013mya} is of this type. 

In the case of the hilltop potential (for example, the top-hat potential considered here), there is a finite upper bound $\Delta$ on the size of the field fluctuation during eternal inflation. This means that, for any particular world line, inflation will always end eventually, even though the amount of inflation may be extremely large. If the average lifetime calculated along an ensemble of world lines is small enough, then eternal inflation will be quenched, {\it i.e.} the volume of non-inflating regions will increase exponentially quickly relative to the volume of inflating regions. This brings up the question: for inflation models which happen at very low energy scales, so that $V_0$ and $\Delta$ are very small in Planck units, are there circumstances under which eternal inflation will be quenched? In the next section, we consider generic hilltop inflationary potentials, and show that slow-roll inflation is in fact always eternal, even when inflation occurs at arbitrarily low energy scale.

\section{Hilltop eternal inflation}
\label{sec:Hilltop}

\begin{figure}
\begin{center}
\includegraphics[width=0.45\textwidth]{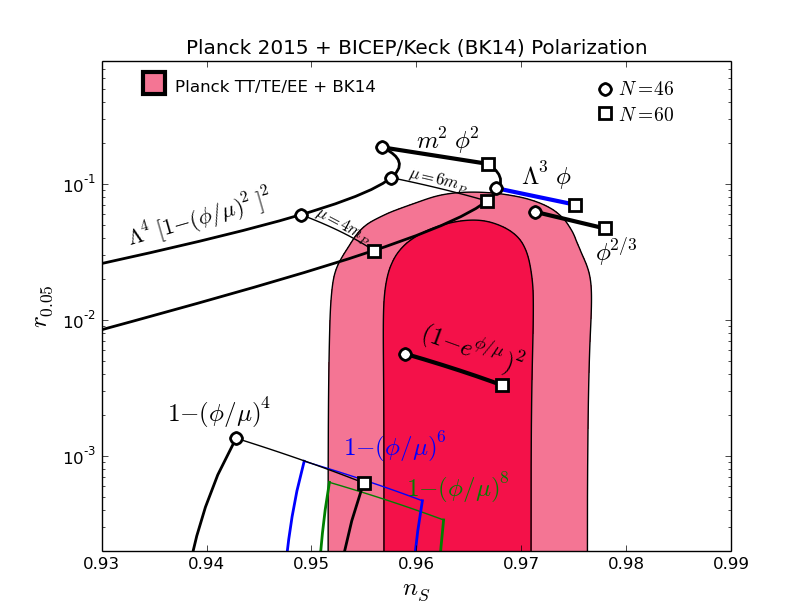}
\end{center}
\caption{Generic inflation models compared with allowed $68\%$ (dark shaded) and $95\%$ (light shaded) regions from Planck and BICEP/Keck 2014 data. \cite{Ade:2015lrj,Array:2015xqh}}
\label{fig:Plancklimits}
\end{figure}

In this section, we consider more realistic potentials of the ``hilltop'' form, where inflation occurs as the field rolls off a local maximum of the potential, with $V'\left(\phi_0\right) = 0$. Taking $\phi_0 = 0$ and expanding in powers of the field $\phi$, such a potential can always be approximated close to the maximum as
\begin{equation} \label{eq:V-hilltop}
V\left(\phi\right)  = V_0\left[1 - \left(\frac{\phi}{\mu}\right)^p + \cdots\right],
\end{equation}
where the lowest-order operator $\phi^p$ with $p>0$ dominates for $\phi \ll \mu$. Inflation models with $p = 2$, such as the simplest versions of Natural Inflation \cite{Freese:1990rb} typically require $\Delta\phi \sim M_{\rm P}$.  
Potentials with $p > 2$ support ``small-field'' inflation, where the field variation is much less than a Planck mass, $\Delta\phi \ll M_{\rm P}$. \cite{Turner:1993su,Kinney:1995ki,Kinney:1995cc}. 
\footnote{``Saddle-point'' inflation models with odd exponent $p$ can avoid eternal inflation through an appropriate choice of parameters \cite{Kinney:2014jya}, so we consider even powers of $p$.}
For very small field variation $\Delta\phi \ll M_{\rm P}$, the potential becomes ``narrow'' in Planck units, and it is reasonable to ask if, at some scale, the condition $(\ref{eq:EIcondition})$ is violated, and eternal inflation is quenched. We show below that $(\ref{eq:EIcondition})$ holds at all scales, and eternal inflation is supported even when inflation happens at arbitrarily low energy scale. Fig \ref{fig:Plancklimits} shows various hilltop inflation models in comparison with the most recent CMB data. 

\subsection{Case 1: $p = 2$}

Consider first the case $p = 2$. The scalar spectral index is given by
\begin{equation}
n - 1 = -6 \epsilon + 2 \eta,
\end{equation}
where
\begin{equation}
\epsilon = \frac{M_{\rm P}^2}{2} \left(\frac{V'}{V}\right)^2 \simeq 2 \left(\frac{M_{\rm P}}{\mu}\right)^2 \left(\frac{\phi}{\mu}\right)^2,
\end{equation}
and 
\begin{equation} \label{eq:etaquad}
\eta = M_{\rm P}^2 \frac{V''}{V} = -2 \left(\frac{M_{\rm P}}{\mu}\right)^2.
\end{equation}
Since $\epsilon \rightarrow 0$ when $\phi \ll \mu$, we have 
\begin{equation}
\eta = {\rm const.},\ \epsilon \ll \eta,
\end{equation}
and 
\begin{equation}
1 - n \rightarrow -2 \eta = 4 \left(\frac{M_{\rm P}}{\mu}\right)^2 = {\rm const.},
\end{equation}
and the Planck constraint $1 - n < 0.05$ implies $\mu > M_{\rm P}$. Eternal inflation occurs when $P\left(k\right) > 1$, where
\begin{equation}
P\left(k_{EI}\right) = \frac{\mu^2 H^2}{16 \pi^2 M_{\rm P}^4} \left(\frac{\mu}{\phi_{EI}}\right)^2 = 1,
\end{equation}
so that, taking $\Delta \sim \phi_{EI}$,
\begin{equation}
\frac{\Delta}{M_{\rm P}} = \frac{1}{4 \pi} \left(\frac{\mu^2 H}{M_{\rm P}^3}\right).
\end{equation}
From (\ref{eq:V-hilltop}) with $\phi \ll \mu$, 
\begin{equation}
H^2 \simeq \frac{V_0}{3 M_{\rm P}^2},
\end{equation}
so that
\begin{equation}
\frac{\Delta}{M_{\rm P}} = \frac{1}{4 \sqrt{3} \pi} \left(\frac{\mu^2 \sqrt{V_0}}{M_{\rm P}^4}\right).
\end{equation}
The condition (\ref{eq:EIcondition}) for eternal inflation is then
\begin{equation}
\left(\frac{\mu}{M_{\rm P}}\right)^2 > \frac{2}{\sqrt{3}},
\end{equation}
or, using Eq. (\ref{eq:etaquad}),
\begin{equation}
\left\vert\eta\right\vert < \sqrt{3}.
\end{equation}
Therefore, eternal inflation is always consistent for $p = 2$. 

\subsection{Case 2: $p > 2$}.

We now consider the case $p > 2$. Inflation ends when the first slow-roll parameter exceeds unity,
\begin{equation}
\epsilon = \frac{M_{\rm P}^2}{2} \left(\frac{V'\left(\phi\right)}{V\left(\phi\right)}\right)^2 \simeq \frac{p^2 M_{\rm P}^2}{2 \mu^2} \left(\frac{\phi}{\mu}\right)^{2 \left(p - 1\right)},
\end{equation}
so that $\epsilon\left(\phi_e\right) = 1$ gives
\begin{equation}
\frac{\phi_e}{\mu} = \left[\frac{2}{p^2}\left(\frac{\mu}{M_{\rm P}}\right)^2\right]^{1 / 2 \left(p -1\right)}. 
\end{equation}
The number of e-folds as a function of field value is, for $p > 2$,
\begin{equation}
N = \frac{1}{\sqrt{2} M_{\rm P}} \int_{\phi_e}^{\phi_N}{\frac{d\phi}{\sqrt{\epsilon}}} \simeq \frac{1}{p \left(p - 2\right)} \left(\frac{\mu}{M_{\rm P}}\right)^2 \left(\frac{\mu}{\phi_N}\right)^{p - 2},
\end{equation}
which is {\it independent} of $\phi_e$ for $\mu \ll M_{\rm P}$  \cite{Kinney:1995ki,Kinney:1995cc}. We then have
\begin{equation} \label{eq:phiN}
\frac{\phi_N}{\mu} = \left[\frac{1}{N p \left(p - 2\right)} \left(\frac{\mu}{M_{\rm P}}\right)^2\right]^{1/\left(p -2\right)}.
\end{equation}
The curvature perturbation amplitude is
\begin{equation} \label{eq:PSnorm}
P\left(k_*\right) = \frac{H^2}{8 \pi^2 M_{\rm P}^2 \epsilon\left(\phi_*\right)} \simeq 2.2 \times 10^{-9},
\end{equation}
where the latter number is the normalization obtained from the Planck observation of the Cosmic Microwave Background (CMB) perturbation spectrum \cite{Ade:2015xua}. The scalar spectral index is given by
\begin{equation} \label{eq:ns}
n - 1 = - 6 \epsilon_* + 2 \eta_* \simeq 2 \eta_* \simeq - \left(\frac{p-1}{p-2}\right) \frac{2}{N_*}.
\end{equation}

\begin{figure}[h]
\begin{center}
\includegraphics[width=0.45\textwidth]{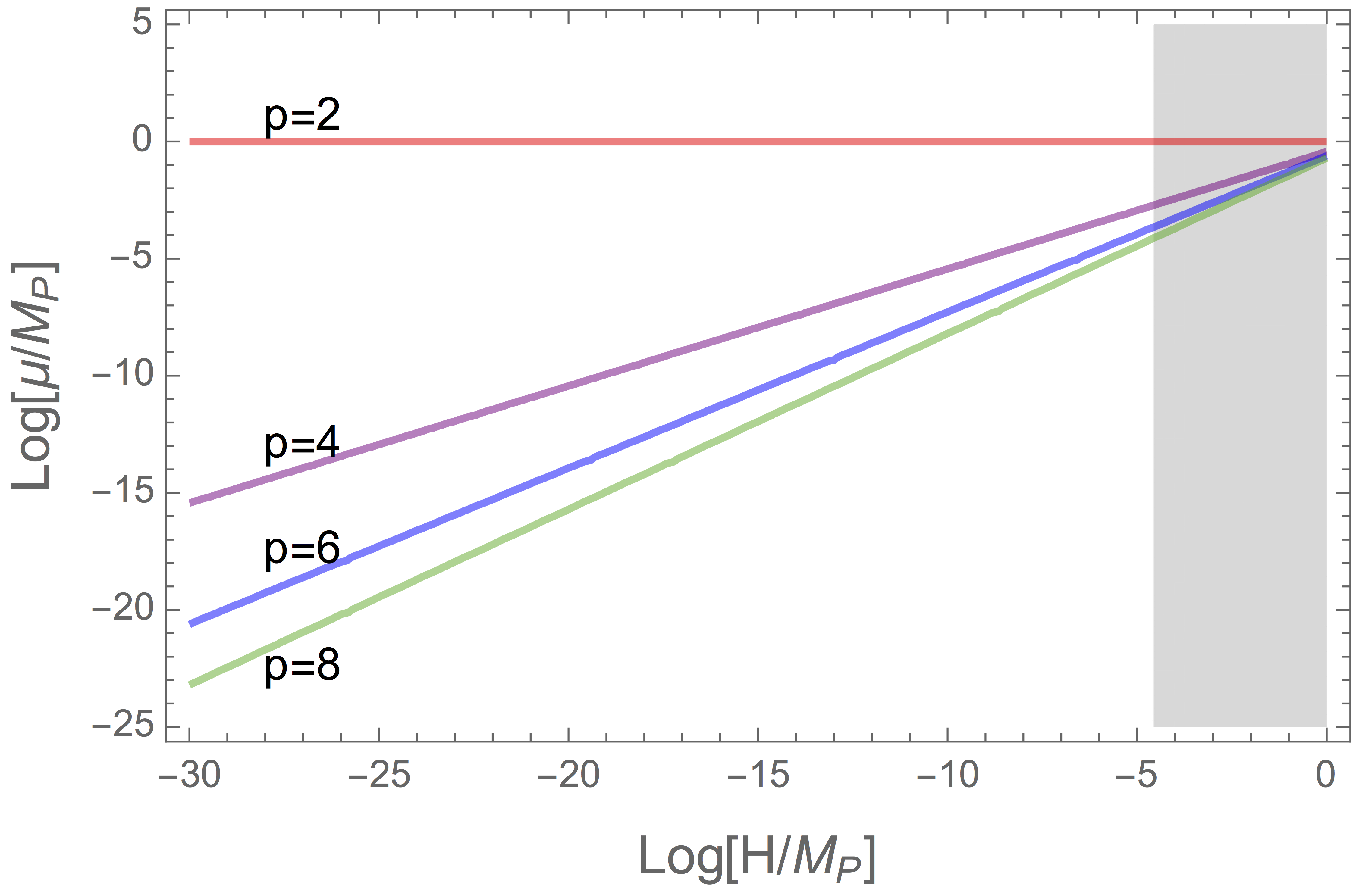}
\caption{Contours satisfying $\Delta/\langle \delta \phi_Q \rangle = 1/\sqrt{3}$ with $(\ref{eq:smallfieldEI})$ for $p=(2,4,6,8)$ from top to bottom, respectively. In the region above each line, eternal inflation is supported. The gray region is excluded from the Planck upper bound on the tensor-to-scalar ratio.}
\label{fig:H-mu}
\end{center}
\end{figure}

Eternal inflation occurs when the amplitude of quantum fluctuations is comparable to the classical field variation in a Hubble time,
\begin{equation}
\left\langle \delta\phi\right\rangle_Q = \frac{H}{2 \pi} = \frac{\dot\phi_{\rm Cl}}{H},
\end{equation}
where, in slow roll inflation,
\begin{equation}
\frac{\dot\phi_{\rm Cl}}{H} \simeq \frac{V'\left(\phi\right)}{3 H^2} \simeq \frac{p V_0}{3 \mu H^2}\left(\frac{\phi}{\mu}\right)^{p-1}.
\end{equation}
Therefore, eternal inflation happens for $\phi < \Delta$, where
\begin{equation}
\frac{\Delta}{\mu} = \left(\frac{3 \mu H^3}{2 \pi p V_0}\right)^{1/\left(p-1\right)} = \left(\frac{\mu H}{2 \pi p M_{\rm P}^2}\right)^{1/\left(p-1\right)}.
\end{equation}
The condition (\ref{eq:EIcondition}) for eternal inflation is then
\begin{eqnarray} \label{eq:smallfieldEI}
\frac{\Delta}{\left\langle\delta\phi\right\rangle_Q} &=& \left(2 \pi\right) \left[\frac{1}{2 \pi p} \left(\frac{\mu}{M_{\rm P}}\right)^p \left(\frac{H}{M_{\rm P}}\right)^{2 -p}\right]^{1/\left(p-1\right)}\cr &>& \frac{1}{\sqrt{3}}.
\end{eqnarray}
The parameter space of eternal inflation in the $(H, \mu)$-plane can be found as shown in Fig.~\ref{fig:H-mu}. 
%
%

From Eqs. (\ref{eq:phiN},\ref{eq:PSnorm}), we can write the Hubble parameter during inflation in terms of the observed power spectrum normalization as:
\begin{eqnarray}
&&\left(\frac{H}{M_{\rm P}}\right)^2 = 4 \pi^2 p^2 P\left(k_*\right)\cr
&&\times \left(\frac{1}{N_* p \left(p - 2\right)}\right)^{2 \left(p - 1\right) / \left(p - 2\right)} \left(\frac{\mu}{M_{\rm P}}\right)^{2p/\left(p-2\right)},
\end{eqnarray}
and the condition (\ref{eq:smallfieldEI}) reduces to the very simple form:
\begin{eqnarray} 
\frac{\Delta}{\langle \delta \phi \rangle_Q} &=& \left[N_* \left(p - 2\right)\right] \left[P\left(k_*\right)\right]^{\left(2 - p\right) / 2 \left(p - 1\right)}\cr
 &=& \left[\frac{2 \left(p-1\right)}{1-n}\right] \left[P\left(k_*\right)\right]^{\left(2 - p\right) / 2 \left(p - 1\right)} > \frac{1}{3},
\end{eqnarray}
which is trivially satisfied for $P\left(k_*\right) \sim 10^{-9}$, $1 -n \simeq 0.04$, and $p > 2$. This is the main result of this paper, and we therefore conclude that in Hilltop inflation scenarios, eternal inflation  is supported \textit{irrespective of the energy scale of inflation}.

\section{Conclusions}
\label{sec:Conclusions}

In this paper, we have considered several scenarios for eternal inflation, which occurs when the quantum fluctuations of the inflaton field dominate over the classical field evolution,
\begin{equation}
\frac{\left\langle \delta\phi\right\rangle_Q}{\delta\phi_{Cl}} = \frac{H^2}{2 \pi \dot\phi} > 1. 
\end{equation}
A phase of eternal inflation can occur for two distinct dynamical reasons: (1) The expansion rate $H$ becomes large, or (2) the field variation $\dot\phi$ becomes small, or a combination of those two. In chaotic inflation models, with $V\left(\phi\right) \propto \phi^p$ for some power $p \geq 1$, eternal inflation is characterized by $H_{EI} \gg H_*$, where $H_*$ is the expansion rate during classical slow-roll inflation. Conversely, in hilltop inflation models, $V\left(\phi\right) \propto 1 - \left(\phi/\mu\right)^p$, in which the inflaton rolls off a local maximum of the potential, $H_{EI} \sim H_*$: the expansion rate during eternal inflation is nearly identical to that during slow-roll inflation. Furthermore, unlike chaotic inflation models, the region of the potential supporting eternal inflation has {\it finite} width $\Delta$ in hilltop models. This means that a phase of eternal inflation, measured in any particular Hubble volume, has a finite lifetime, with expectation value
\begin{equation}
\left\langle t\right\rangle \simeq \frac{4 \pi^2 \Delta^2}{H^3}.
\end{equation}
However, if we consider a uniform ensemble of initial Hubble patches, any given Hubble patch will increase in volume exponentially with time,
\begin{equation}
{\mathcal V} \propto e^{3 H t}.
\end{equation}
Therefore, if the expected lifetime of inflation in a Hubble patch is large in units of a Hubble time,
\begin{equation}
\left\langle t\right\rangle > 1 / \left(3 H\right),
\end{equation}
then at any finite time $t$, the volume of the universe will be exponentially dominated by Hubble patches still undergoing inflation, and inflation can said to be truly eternal. The lifetime depends on the width in field space of the region on the potential which supports eternal inflation, which in turn depends on the scale of inflation. 

In this paper, we considered hilltop inflation models consistent with current data, and showed that eternal inflation can occur {\it regardless} of the energy scale of inflation. This is perhaps not surprising, since it is at base a statement about the flatness of the inflaton potential: if the peak in $V\left(\phi\right)$ is sufficiently flat, then the lifetime of a quantum-dominated state is long enough to support eternal inflation. However, the conclusion is counterintuitive from the standpoint of decoupling, since it means that quantum processes can have a large influence on the cosmological spacetime, generating order-unity comoving curvature perturbations, even when all of the energy scales in the field Lagrangian are far below the scale of quantum gravity. Quantum effects can play a critical role in determining the global structure of the cosmological spacetime, for example, for inflation at energy scales of order a ${\rm TeV}$. Of course, this is a statement about whether or not eternal inflation {\it can} happen, not necessarily about whether it {\it will}: the latter is a question of boundary conditions. If the field is not fully relaxed to the inflationary attractor, for example, it is easily possible that $\dot\phi \gg \dot\phi_{SR}$, and classical field evolution dominates over quantum even near $\phi = 0$ \cite{Tzirakis:2007bf}. Without a definite understanding of the initial conditions for inflation, it is not possible to arrive at a definite answer to the question of whether or not eternal inflation, in fact, was triggered in the early universe.

\section*{Acknowledgments}

GB and WIP acknowledge support from the MEC and FEDER (EC) Grants SEV-2014-0398 and FPA2014-54459 and the Generalitat Valenciana under grant PROME- TEOII/2013/017.
GB acknowledges partial support from the European Union FP7 ITN INVISIBLES MSCA PITN-GA-2011-289442 and InvisiblesPlus (RISE) H2020-MSCA-RISE-2015-690575.
WHK thanks the University of Valencia, where this work was completed, for generous hospitality and support, and thanks Paul Steinhardt and Katherine Freese for helpful conversations. WHK is supported by the National Science Foundation under grant NSF-PHY-1417317. This work was performed in part at the University at Buffalo Center for Computational Research.

\bibliography{paper}

\end{document}